\documentclass[a4paper,11pt]{article}

\usepackage{pos}
\usepackage{wrapfig}
\usepackage{subcaption}
\usepackage{cleveref}

\title{2D Convolutional Neural Network for Event Reconstruction in IceCube DeepCore}

\ShortTitle{2D Convolutional Neural Network for Event Reconstruction in IceCube DeepCore}

\author{The IceCube Collaboration \\{\normalsize \normalfont(a complete list of authors can be found at the end of the proceedings)}\\}

\emailAdd{josh.peterson@icecube.wisc.edu}
\emailAdd{mvprado@icecube.wisc.edu}
\emailAdd{kael.hanson@wisc.edu}

\abstract{
 
IceCube DeepCore is an extension of the IceCube Neutrino Observatory designed to measure GeV scale atmospheric neutrino interactions for the purpose of neutrino oscillation studies.  Distinguishing muon neutrinos from other flavors and reconstructing inelasticity are especially difficult tasks at GeV scale energies in IceCube DeepCore due to sparse instrumentation.  Convolutional neural networks (CNNs) have been found to have better success at neutrino event reconstruction than conventional likelihood-based methods.  In this contribution, we present a new CNN model that exploits time and depth translational symmetry in IceCube DeepCore data and present the model’s performance, specifically for flavor identification and inelasticity reconstruction.

\vspace{4mm}
{\bfseries Corresponding authors:}
J.H. Peterson$^{1*}$, M. Prado Rodriguez$^{1}$, K. Hanson$^{1}$\\
{$^{1}$ \itshape University of Wisconsin - Madison}\\[4mm]
$^*$ Presenter

\ConferenceLogo{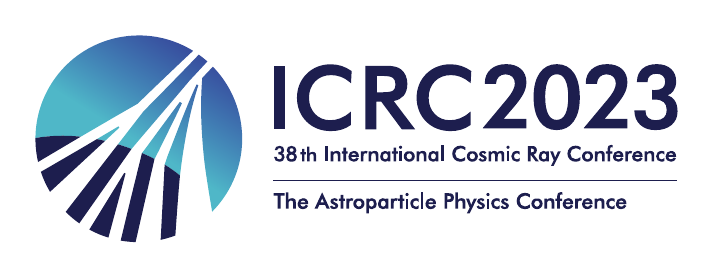}

\FullConference{The 38th International Cosmic Ray Conference (ICRC2023)\\ 26 July -- 3 August, 2023\\ Nagoya, Japan}
}

\begin{document}

\maketitle

\section{Introduction}\label{sec1}

The IceCube Neutrino Observatory is a neutrino detector located at the South Pole.  When high energy neutrinos interact with the Antarctic ice they produce fast moving charged secondary particles that produce Cherenkov radiation.  The IceCube detector consists of 86 strings of digital optical modules (DOMs) that detect this Cherenkov radiation \cite{Aartsen_2017}. IceCube DeepCore is an infill of the IceCube Neutrino Observatory that utilizes more densely instrumented strings to lower the energy threshold from hundreds of GeV to a few GeV \cite{PhysRevD.99.032007}.  Many of the IceCube experiment's physics goals require reconstructing observables such as neutrino direction, neutrino energy, and neutrino flavor.

When a muon neutrino / antineutrino undergoes a charged current deep inelastic scattering it produces an energetic muon / antimuon that can travel a significant distance and produce a detectable track of light \cite{Aartsen_2014}.  This is referred to as a track event.  At lower energies, for neutral current deep inelastic scattering events, and for charged current deep inelastic scattering of electron neutrinos / antineutrinos and most tau neutrinos / antineutrinos, only the hadronic cascade will be visible \cite{PhysRevD.99.032007}.  This is referred to as a cascade event.  Thus, muon neutrinos / antineutrinos can be seperated from the other flavors using the IceCube detector.

The inelasticity is the fraction of neutrino energy that is deposited in the hadronic cascade of a deep inelastic scattering.  The average inelasticity for muon neutrinos and muon antineutrinos is different \cite{PhysRevD.99.032004, Devi_2014}.  This can allow us to separate neutrinos from antineutrinos statistically using the IceCube detector.

Both neutrino flavor and inelasticity are observables that require evaluating the morphology of individual events.  For particle identification (PID), we look for the existence of an outgoing muon / antimuon track \cite{Aartsen_2014}.  At the energies relevant for DeepCore, the energetic muon / antimuon is a minimum ionizing particle, and thus the average length the muon / antimuon  travels scales linearly with the energy of the muon / antimuon.  Thus we can look at the length of the outgoing track and the total light deposited in the hadronic cascade to determine the inelasticity \cite{Aartsen_2014}.  The density of the DOMs in the ice means that events at lower energies have lower resolutions, making standard reconstruction techniques more difficult to use.  It has been shown that machine learning based reconstruction methods are faster and more precise than the photon table based reconstruction methods that have been previously used in IceCube \cite{GNN}.  In this paper we describe a new convolutional neural network (CNN) developed to measure PID and inelasticity in IceCube DeepCore.

\section{Methods}

\subsection{Monte Carlo Data and Selection} \label{select}

For training and testing the CNN we use simulated data that includes the following restrictions:

\begin{itemize}
\item The energy must be between 3 GeV and 1 TeV for inelasticity reconstruction, 5 GeV to 1 TeV for PID reconstruction.

\item There must be at least eight photon hits and at most 250 photon hits in the event.

\item The interaction must occur between 2106 meters and 2450 meters deep in the ice in a circle of radius 100 meters surrounding the center of DeepCore.
\end{itemize}

We elect to use relatively loose selection criteria so that we have more events to train the CNN with and so that in the future we can apply the trained CNN to existing data sets that have stricter selection criteria.  The energy threshold was lowered from 5 GeV to 3 GeV for inelasticity to match existing matter dependant studies.  For PID reconstruction, we need to select a sample that contains both tracks and cascades, so we take muon and electron neutrino / antineutrino events, including both neutral and charged current interactions.  For inelasticity reconstruction we just need to focus on track events, so we only take muon neutrino / antineutrino charged current interactions.  After imposing these restrictions our data has the composition shown in Table \ref{compo}.

\begin{table}[h]
\centering
\caption{Composition of Monte Carlo data used for training the CNN.  A track event is a muon neutrino / antineutrino CC interaction.  Everything else is labelled a cascade event.}
\begin{tabular}{ |c||c|c| } 
 \hline
  & tracks & cascades \\ 
 \hline \hline
 PID & 4,495,634 & 2,835,831 \\ 
 \hline
 inelasticity & 4,827,397 & 0 \\
 \hline
\end{tabular}
   \label{compo}
\end{table}

We then split the data into three sets: $4\%$ for a testing the CNN, $10\%$ for a validating performance during training, and the rest is used as a training set.

\subsection{Monte Carlo Data Preparation}

\begin{figure}[h]
  \centering
  \begin{minipage}[b]{0.45\textwidth}
   \includegraphics[width=\textwidth]{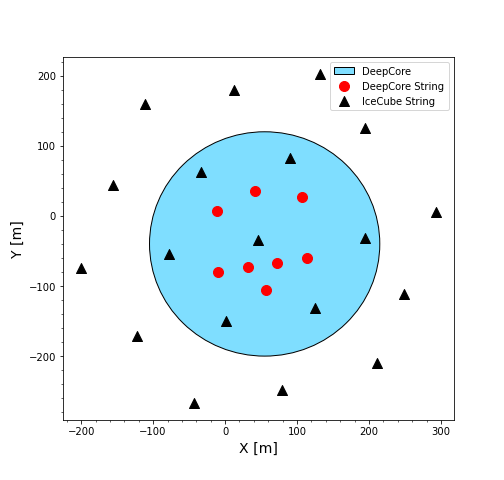}
   \caption{Spacing of strings in the x-y plane of the DeepCore detector.  The black triangles represent strings of DOMs identical to the strings that are found in the rest of IceCube.  The red spots represent DeepCore strings that were installed specifically for the DeepCore detector and are more densly instrumented.  All of the strings that are within the blue circle are used in the DeepCore detector.}
   \label{DeepCore}
  \end{minipage}
  \hfill
  \begin{minipage}[b]{0.45\textwidth}
   \includegraphics[width=\textwidth]{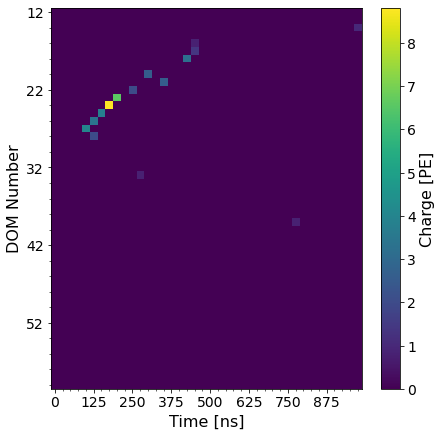}
   \caption{Example of a string image.  Each row is a DOM, and the rows are ordered based on depth in the ice.  The pixels are populated with the total charge that each DOM saw starting at a particular slice of time.  Each event will have an image for every string in the DeepCore detector.}
   \label{string_example}
  \end{minipage}
\end{figure}

Our data should have translational symmetry in both space and time.  However, the strings in the DeepCore detector are non-uniformly spaced in the x-y plane, as shown in Figure \ref{DeepCore}, so exploiting the x-y translational symmetry cannot be easily done.  The DOM spacing is uniform along the strings, so we can exploit the translational symmetry along depth and along time.  Thus, for every event we compose a 2-dimensional image for each string in the DeepCore detector.  The images consist of the DOMs ordered by depth on one axis and time sliced into 25 nanosecond time segments on the other axis.  The charge of the photon hits seen by a particular DOM in an event is summed, and the pixel containing the earliest photon hit in a particular DOM is populated with the total charge.  An example of such a string image is shown in Figure \ref{string_example}.

There are two types of strings of DOMs in the DeepCore detector.  Seven of the strings are identical to the strings that are used in the rest of the IceCube detector.  We will refer to those strings as IceCube strings.  There are eight strings that are more densely instrumented surrounding a single IceCube string.  We will refer to those strings as DeepCore strings.  We utilize all of the strings in the blue circle in Figure \ref{DeepCore} in our CNN.  For the inelasticity CNN we also incorporate the IceCube strings surrounding the DeepCore detector shown in Figure \ref{DeepCore}.  This is done because it was found that we obtain better reconstruction results at higher energy.  We will do the same for PID in the future, but the results shown in this paper for PID do not include those additional IceCube strings.

We elect to only use DOMs that are below a particular dust layer at about 2000 meters to 2100 meters deep that reduces the optical quality of the ice \cite{Ice_paper}.  This creates the complication that the string images constructed from DeepCore strings will have more rows than the images constructed from IceCube strings since there are more DOMs below the dust layer on the DeepCore strings than on the IceCube strings \cite{Aartsen_2017}.  Thus we pad the IceCube string images with empty DOM rows in such a way that IceCube string DOMs and DeepCore string DOMs with similar depths will be in the same row in the string images.  We then stack the DeepCore string images and padded IceCube string images together to feed to the CNN.

\subsection{Neural Network Architecture and Training}

\begin{figure}[h]
\includegraphics[width=\textwidth]{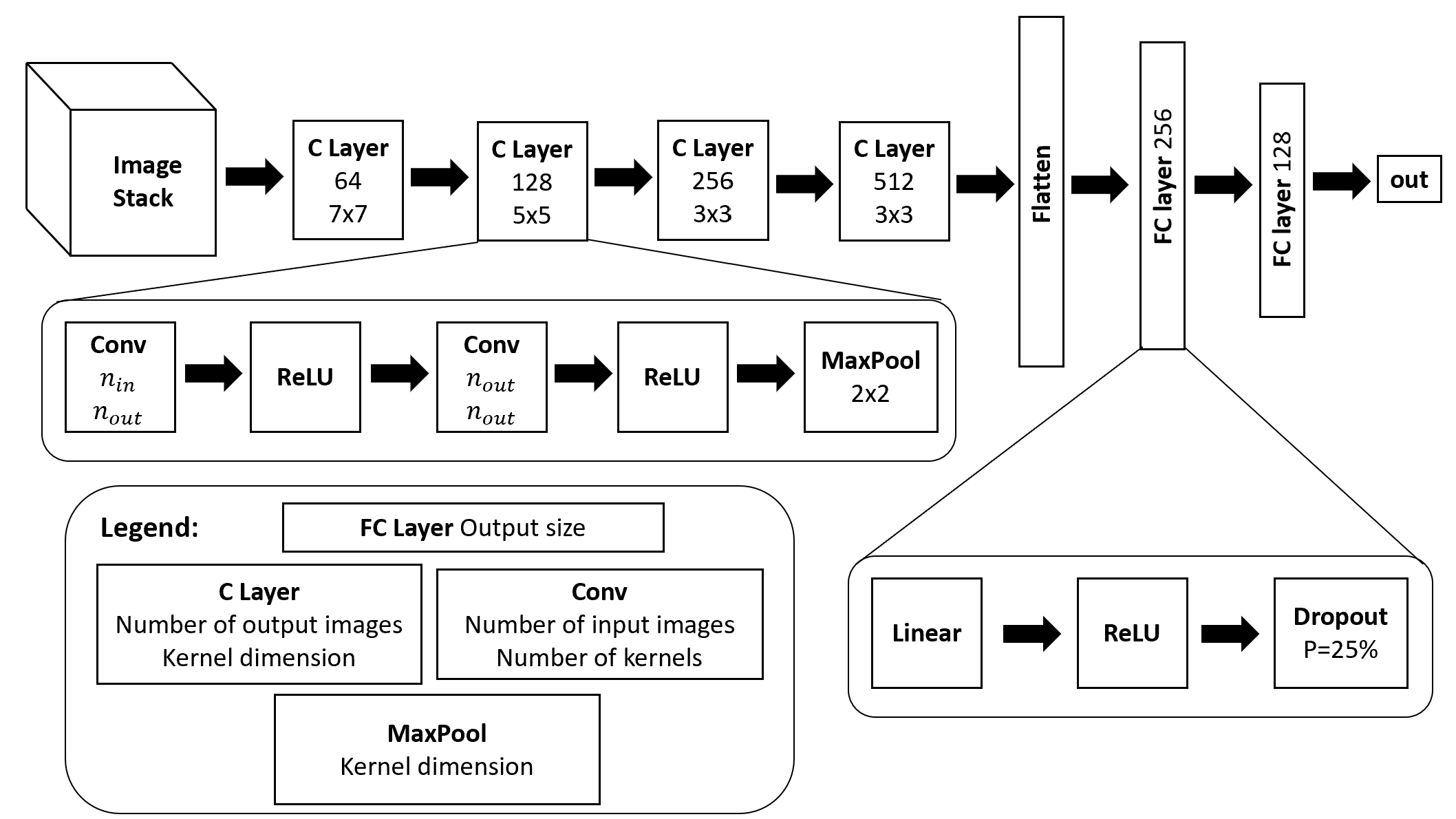}
\caption{Architecture of our 2D CNN.  The "out" at the the end represents whatever output one might need, for this paper it would be inelasticity or a track or cascade confidence.}
\label{cnn_diagram}
\end{figure}

Our CNN includes four convolutional blocks followed by a two layer multi-layer perceptron (MLP).  Each convolution block includes two 2-dimensional convolution layers with rectified liner unit (ReLU) activation functions.  Then we apply a max pooling layer to the output of the convolutional layers.  For the MLP we use a dropout rate of $25\%$ to help regularize the CNN. A diagram of the neural network architecture is shown in Figure \ref{cnn_diagram}.

For the inelasticity CNN, we have a single output that is passed through a sigmoid activation function after the MLP.  Inelasticity is bounded between 0 and 1, so the sigmoid activation function insures that the output of the CNN upholds these bounds.

For PID, we have two outputs: a label for the presence of a significant muon track and a label for a significant hadronic cascade.  We use these labels to represent the fact that some events are dominated by the hadronic cascade, some are dominated by the muon track, and some events have a significant cascade and track.  We use true inelasticity as a metric for when we identify an event as having a significant track and cascade.  Table \ref{labels} shows how each event is assigned its labels.  The cuts in the table are determined by training many networks with different cuts and picking the configuration with the best performance.

\begin{table}[h]
\centering
\caption{The conditions used to label each event.  Each event has a muon track and hadronic cascade label to indicate if there is a visible muon track or hadronic cascade.}
\begin{tabular}{ |c|c||c|c| } 
 \hline
 \multicolumn{2}{|c||}{Muon Track label} & \multicolumn{2}{|c|}{Hadronic Cascade label} \\
 \hline \hline
  label & condition & label & condition \\ 
 \hline
 track & $\nu_\mu$ CC events & cascade & $\nu_\mu$ NC events, $\nu_e$ events, $\nu_\mu$ CC events w/ $y>0.8$ \\ 
 \hline
 no track & $\nu_\mu$ NC events, $\nu_e$ events & no cascade & $\nu_\mu$ CC events w/ $y<0.8$ \\ 
 \hline
\end{tabular}
\label{labels}
\end{table}

For training, we use the Adam algorithm \cite{kingma2017adam} for optimization with a learning rate of 0.0001.  We stop the training when the validation loss achieves a minimum and starts to increase upon following epochs.

For the inelasticity CNN, we use the following L1 loss:
\begin{equation} \label{y_loss}
Loss_{inelasticity} = |y_{reconstructed} - y_{true}|.
\end{equation}
This choice was made because we find that we achieve better performance at higher energies with this loss compared to L2 loss.  For the PID CNN we use the binary cross entropy (BCE) loss applied to both outputs:
\begin{equation} \label{pid_loss}
Loss_{PID} = 0.7\times BCE(output_{track}; label_{track}) + BCE(output_{cascade}; label_{cascade}).
\end{equation}
The binary cross entropy loss for the track label is weighted by a factor of 0.7, which was found to improve performance via a grid scan.

\section{Results and Discussion}

\subsection{Particle Identification}

We find that the track and cascade labels provide very similar outputs, and so we only show here the output of the track label.  However, we find that using this labelling scheme and loss gives better results than just using the cross entropy loss, which is why we keep it regardless of the unintended behavior.

\begin{figure}[h]
  \centering
  \begin{minipage}[b]{0.45\textwidth}
   \includegraphics[width=\linewidth]{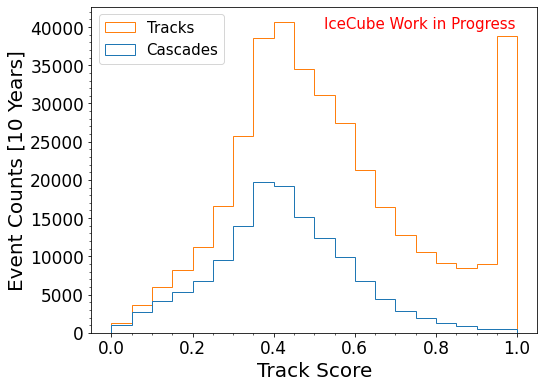}
   \caption{Histogram of MC events with all three flavors of neutrinos binned by PID score.}
   \label{pid_distro}
  \end{minipage}
  \hfill
  \begin{minipage}[b]{0.45\textwidth}
   \includegraphics[width=\linewidth]{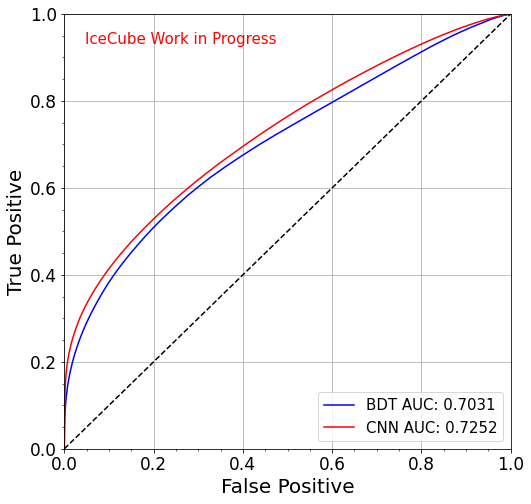}
   \caption{Receiving operating characteristic curve for the 2D CNN evaluated on a set of MC events with all three flavors of neutrinos.  The 2D CNN classifies better than a BDT trained for the same task.}
   \label{roc}
  \end{minipage}
\end{figure}

 For Figure \ref{pid_distro} and Figure \ref{roc}, we use all muon, electron, and tau neutrino / antineutrino Monte Carlo (MC) events that pass the data selection criteria outlined in Section \ref{select}.  After checking the normalized PID distributions for the test set and the training set, we find they are nearly identical, suggesting that overfitting has been well-regulated.

Figure \ref{pid_distro} is the distribution of PID outputs for the set of MC events described previously.  There are three notable features of this distribution; the spike in track events near a PID score of 1, The collection of events around a PID score of 0.5, and the cascade rich tail below a PID score of 0.2.  The spike in track events near a PID score of 1 and the central collection of cascade and track events are common features for other PID reconstruction algorithms for the DeepCore detector \cite{LeonardDeHolton:2022lR, icecubecollaboration2023measurement}.   The collection of cascade and track events around a PID score of 0.4 to 0.5 and the lack of a spike of cascade events around a PID score of 0 are a reflection of the fact that at low energies many track events look like cascades, and only when there is a definite muon track can we be confident in what a particle's identity is.  The cascade rich tail is a feature that isn't seen in similar reconstruction methods, and suggests that the CNN has started to find features in the data unique to cascade events.

Figure \ref{roc} shows the receiver operating characteristic curves for our CNN and a boosted decision tree (BDT) that is used in another analysis \cite{LeonardDeHolton:2022lR}. This plot demonstrates that the CNN method is better at identifying tracks and cascades than the BDT based method, which relies on high level reconstructed inputs.

\subsection{Inelasticity}

\begin{figure}[t]
\subfloat[]{\includegraphics[width = 0.475\textwidth]{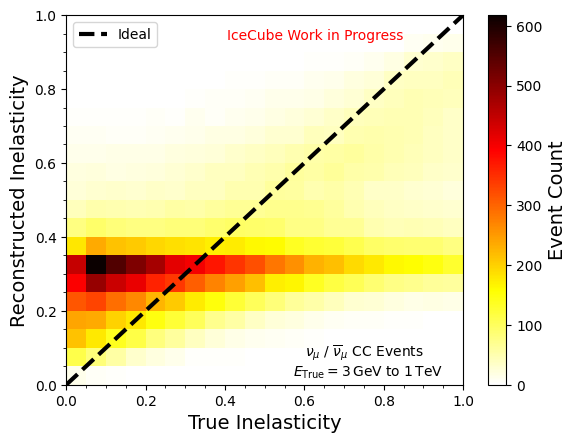}\label{full_e} }
\subfloat[]{\includegraphics[width = 0.475\textwidth]{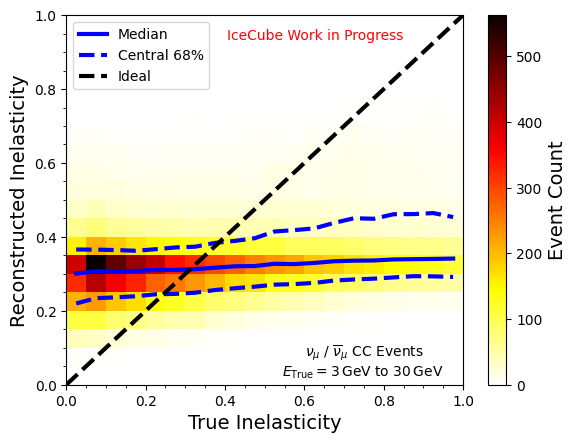}\label{low_e}} \\
\subfloat[]{\includegraphics[width = 0.475\textwidth]{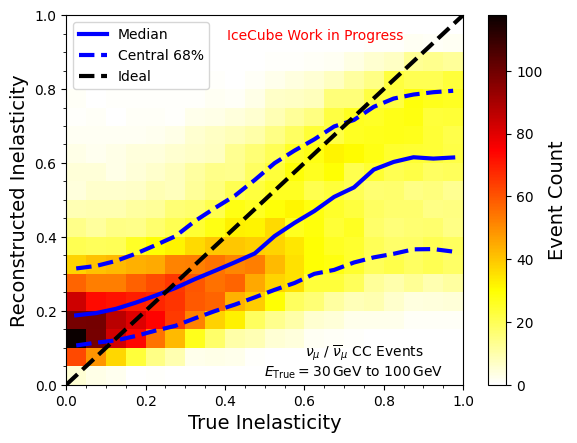}\label{mid_e}} 
\subfloat[]{\includegraphics[width = 0.475\textwidth]{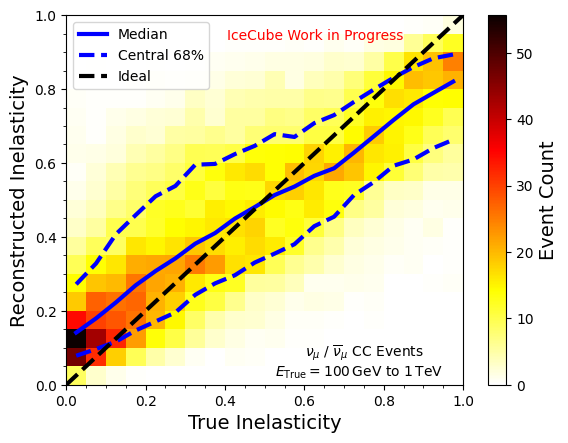}\label{high_e}} 
\caption{MC events binned by true inelasticity and inelasticity reconstructed by the CNN.  The dashed black line indicates where events would be assuming a perfect reconstruction.  Plot (a) contains all events in the test set, whereas (b), (c), and (d) show events separated by different true energy ranges.}
\label{some example}
\end{figure}

Figure \ref{full_e} is a two dimensional histogram of events binned by the true inelasticity and the inelasticity from CNN reconstruction.  An ideal reconstruction would have all events in the bins along the dashed black line.  There are two populations of events that can be seen here.  The flat distribution are events that are hard to reconstruct, and so the network learns to assign roughly the mean inelasticity of those events.  The second population of events can be seen faintly in the upper-right corner of the plot.  These are events that are being reconstructed more effectively.

A good proxy for the difficulty of reconstruction is the energy of the event.  Lower energy events will have less photon hits and hence lower resolution.  Thus, we can make cuts in the energy to separate the two populations.

Figure \ref{low_e}, Figure \ref{mid_e}, and Figure \ref{high_e} are similar histograms to Figure \ref{full_e} but with the events split into three populations based on energy (the specific energy ranges are shown in the plots).  Figure \ref{low_e} clearly shows the first population of events without the second.  There is a slight asymmetry in the reconstructed inelasticity distribution which can be used to slightly separate neutrinos from antineutrinos.  To quantify this separation power, we separate events with true energies of 3 GeV to 20 GeV into two bins, $y_{reco}>0.32$ and $y_{reco}<0.32$. Then we calculate the ratio of antineutrino events to all events in each bin and we find ratios of 0.34 and 0.31 respectively.

Figure \ref{mid_e} shows a definite improvement in reconstruction when compared to Figure \ref{low_e}, and Figure \ref{high_e} even better reconstruction than Figure \ref{mid_e}.  This is expected because the larger the energy of the event, the more DOMs will detect photons and thus it will be easier to resolve the event morphology.  The second plot shows that we can push the inelasticity reconstruction down to roughly 30 GeV.

\section{Conclusion}\label{sec3}

We have developed a new 2D CNN architecture as well as a new method for preparing IceCube DeepCore events that is effective for reconstruction tasks in IceCube DeepCore.  We also show that when this CNN is applied to PID reconstruction it outperforms a BDT based method used in other DeepCore analyses, and when we apply the CNN to inelasticity reconstruction we can effectively measure inelasticity down to 30 GeV and gain some neutrino antineutrino separating power below 20 GeV.  It is likely that this CNN will be developed further, perhaps with different loss functions or more optimized architectures.  We will also incorporate the IceCube strings surrounding DeepCore into the PID CNN in the near future.  We then hope to apply this CNN to inelasticity based studies and/or matter effect dependent oscillation studies, where separating neutrinos from anti-neutrinos can greatly benefit sensitivity \cite{Devi_2014}.

\bibliographystyle{ICRC}
\bibliography{references}

\clearpage

\section*{Full Author List: IceCube Collaboration}

\scriptsize
\noindent
R. Abbasi$^{17}$,
M. Ackermann$^{63}$,
J. Adams$^{18}$,
S. K. Agarwalla$^{40,\: 64}$,
J. A. Aguilar$^{12}$,
M. Ahlers$^{22}$,
J.M. Alameddine$^{23}$,
N. M. Amin$^{44}$,
K. Andeen$^{42}$,
G. Anton$^{26}$
% ...
\\
\\
% ...
$^{12}$ Universit{\'e} Libre de Bruxelles, Science Faculty CP230, B-1050 Brussels, Belgium \\
% ...
$^{17}$ Department of Physics, Loyola University Chicago, Chicago, IL 60660, USA \\
$^{18}$ Dept. of Physics and Astronomy, University of Canterbury, Private Bag 4800, Christchurch, New Zealand \\
% ...
$^{22}$ Niels Bohr Institute, University of Copenhagen, DK-2100 Copenhagen, Denmark \\
$^{23}$ Dept. of Physics, TU Dortmund University, D-44221 Dortmund, Germany \\
% ...
$^{26}$ Erlangen Centre for Astroparticle Physics, Friedrich-Alexander-Universit{\"a}t Erlangen-N{\"u}rnberg, D-91058 Erlangen, Germany \\
% ...
$^{40}$ Dept. of Physics and Wisconsin IceCube Particle Astrophysics Center, University of Wisconsin{\textendash}Madison, Madison, WI 53706, USA \\
% ...
$^{42}$ Department of Physics, Marquette University, Milwaukee, WI, 53201, USA \\
% ...
$^{44}$ Bartol Research Institute and Dept. of Physics and Astronomy, University of Delaware, Newark, DE 19716, USA \\
% ...
$^{63}$ Deutsches Elektronen-Synchrotron DESY, Platanenallee 6, 15738 Zeuthen, Germany  \\
$^{64}$ Institute of Physics, Sachivalaya Marg, Sainik School Post, Bhubaneswar 751005, India \\
% ...

\subsection*{Acknowledgements}

\noindent
The authors gratefully acknowledge the support from the following agencies and institutions:
USA {\textendash} U.S. National Science Foundation-Office of Polar Programs,
U.S. National Science Foundation-Physics Division,
U.S. National Science Foundation-EPSCoR,
Wisconsin Alumni Research Foundation,
Center for High Throughput Computing (CHTC) at the University of Wisconsin{\textendash}Madison,
Open Science Grid (OSG),
Advanced Cyberinfrastructure Coordination Ecosystem: Services {\&} Support (ACCESS),
Frontera computing project at the Texas Advanced Computing Center,
U.S. Department of Energy-National Energy Research Scientific Computing Center,
Particle astrophysics research computing center at the University of Maryland,
Institute for Cyber-Enabled Research at Michigan State University,
and Astroparticle physics computational facility at Marquette University;
Belgium {\textendash} Funds for Scientific Research (FRS-FNRS and FWO),
FWO Odysseus and Big Science programmes,
and Belgian Federal Science Policy Office (Belspo);
Germany {\textendash} Bundesministerium f{\"u}r Bildung und Forschung (BMBF),
Deutsche Forschungsgemeinschaft (DFG),
Helmholtz Alliance for Astroparticle Physics (HAP),
Initiative and Networking Fund of the Helmholtz Association,
Deutsches Elektronen Synchrotron (DESY),
and High Performance Computing cluster of the RWTH Aachen;
Sweden {\textendash} Swedish Research Council,
Swedish Polar Research Secretariat,
Swedish National Infrastructure for Computing (SNIC),
and Knut and Alice Wallenberg Foundation;
European Union {\textendash} EGI Advanced Computing for research;
Australia {\textendash} Australian Research Council;
Canada {\textendash} Natural Sciences and Engineering Research Council of Canada,
Calcul Qu{\'e}bec, Compute Ontario, Canada Foundation for Innovation, WestGrid, and Compute Canada;
Denmark {\textendash} Villum Fonden, Carlsberg Foundation, and European Commission;
New Zealand {\textendash} Marsden Fund;
Japan {\textendash} Japan Society for Promotion of Science (JSPS)
and Institute for Global Prominent Research (IGPR) of Chiba University;
Korea {\textendash} National Research Foundation of Korea (NRF);
Switzerland {\textendash} Swiss National Science Foundation (SNSF);
United Kingdom {\textendash} Department of Physics, University of Oxford.

\end{document}